\documentstyle[preprint,revtex,eqsecnum]{aps}

\begin{document}

\preprint{August 1992 \hfill OITS-493}

             \begin{title}
SU(16) grandunification: breaking scales,\\ proton decay
and neutrino magnetic moment
             \end{title}

\author{{\bf N.G. Deshpande, E. Keith, and Palash B.
Pal\cite{newadd}}}

                \begin{instit}
Institute of Theoretical Science, University of Oregon, Eugene, OR
97403, USA
                \end {instit}

                                \begin{abstract}
We give a detailed renormalization group analysis for the SU(16)
grandunified group with  general breaking chains in which quarks and
leptons transform separately at intermediate energies. Our analysis
includes the effects of Higgs bosons. We show that the grandunification
scale could be as low as $\sim 10^{8.5}$ GeV and give examples where
new physics could exist at relatively low energy ($\sim 250$ GeV). We
consider proton decay in this model and show that it is consistent
with a low grandunification scale. We also discuss the possible
generation of a  neutrino magnetic moment in the range of $10^{-11}$
to $10^{-10}\mu_B$ with a very small mass by the breaking of the
embedded SU(2)$_\nu$ symmetry at a low energy.
                                \end{abstract}

\narrowtext
\section{Introduction}
                   Recent treatments of the SU(15) grandunification
group \cite{Adler,su15,Pal91,su15higgs,protondecay} have shown that in the
versions of that model which produce ``ununified models''
\cite{ununified} at intermediate energies, grandunification may be
reached at a relatively low energy and that the lowest intermediate
scale may be within the reach of the Superconducting Super Collider
(SSC). It has also been shown recently that the effects of Higgs
bosons in the renormalization group equations of SU(15) models are
quite large \cite{su15higgs}. In the literature, there is also some
discussion of the gauge group SU(16) \cite{su16}, which has some
desirable features that are not found in SU(15) models. Among these
are seperate gauging of baryon and lepton number \cite{su16} and the
embedding of Voloshin symmetry \cite{Volo,unifiable} SU(2)$_\nu$ which
might play an important role in the solar neutrino puzzle
\cite{Davis,anticorr}. In this paper, we will analyze the
renormalization group equations (RGE's) of the SU(16)
grandunification model with  breaking chains in which quarks and
leptons transform separately at intermediate energies. We will include
the effects of Higgs bosons, which are significant, and use the newest
LEP data for couplings at low energy. We give this analysis in Section
\ref{s:rge}.

In the SU(16) model, all known left-handed fermions of a single
generation together with a left-handed antineutrino transform like the
fundamental representation of the gauge group. i.e.
 \begin{eqnarray}
\Psi_{eL} \equiv \left( \hat{\nu}_e\,   \nu_e \,   e^- e^+\,
u_1\,   u_2\,   u_3\,   d_1\,   d_2\,   d_3\,   \hat{u}_1\,
\hat{u}_2 \,   \hat{u}_3\,   \hat{d}_1\,
\hat{d}_2\,   \hat{d}_3 \right) _L \label{PsiL}
\end{eqnarray}
Here, the hat symbol over a particle's symbol denotes its
antiparticle. Of course, mirror fermions must be introduced to make
the model free of anomalies, but we do not need to discuss them
explicitly here. Our interest is to look for chains with low
unification scale. Existence of such chains is known
\cite{Adler,su15,Pal91,su15higgs} in SU(15) models, for which
it has been shown
that  a low unification scale makes the model free from the monopole
problem \cite{monopole} while being perfectly consistent with the
proton lifetime \cite{protondecay,otherpd}. We discuss proton decay
for our SU(16) model in Section \ref{s:pdk}.

Another reason for examining the SU(16) model is to implement the
suggestion of a previous paper \cite{unifiable} which points out that
this group contains the subgroup SU(4)$_l$, in which the left-handed
leptons including an antineutrino transform as the fundamental
representation. Further, SU(4)$_l$ contains Voloshin symmetry which
allows for the magnetic moment of the neutrino to be in the range of
$10^{-11}$ to $10^{-10}\mu_B$ and at the same time allowing only a
small neutrino mass \cite{unifiable,BarbMoha,others}. Magnetic moment
in  this range might be required \cite{VVO} if the anticorrelation
\cite{anticorr} of solar neutrino flux with sun spot number is
confirmed. We shall discuss the implementation in detail in Section
\ref{s:numu}.

\section{SU(16) breaking scheme and RG analysis}\label{s:rge}
     In Fig\@. \ref{generalbreak},  we show the symmetry breaking
scheme of our model.   The purpose of our scheme is to have the
leptonic sector transform separately from the quark sector at
intermediate energies and to have the leptonic sector SU(4)$_l$ break
in the same manner as in the previous SU(4)$_l$ based model
\cite{unifiable}.   In the figure,  we show the representations of the
Higgs fields used to break symmetries at their indicated mass
scales.   We use the hypothesis of minimal fine-tuning
\cite{finetune},  which allows us to choose the mass scales of
submultiplets of these fields at different scales. We are interested
in scenarios where an intermediate gauge group exists at energies of
the order of 1\,TeV so that we shall have new physics at observable
energies.   We will show in this section that such scenarios are
consistent with renormalization group analysis.

In the one-loop approximation,  couplings $\alpha_i=
g_i^2/4\pi$ evolve as
 \begin{eqnarray}
        {\partial\alpha_i\over\partial\ln M}=-{B_i\over2\pi}\alpha_i^2\,   ,
        \end{eqnarray}
which implies
 \begin{eqnarray}
        \alpha_i^{-1}(M_2)=\alpha_i^{-1}(M_1)-{B_i\over2\pi} \ln {M_1 \over
M_2}\,   .
 \end{eqnarray}
For the standard model couplings $\alpha_{1Y}$, $\alpha_{2L}$,
and $\alpha_{3c}$, we
use the conventional normalization which determines the relations
              \begin{eqnarray}
{\rm Tr}\, (T_iT_j) = 2\delta_{ij} \,, \quad
B_i={11\over3}N-{4\over3}(2n_g)-{1\over6}T(S_i)\, ,
\label{Bsm}
               \end{eqnarray}
where the $T_i$'s are the generators of the SU(16) fundamental
representation.   To simplify the boundary conditions, we normalize
the couplings $\alpha_i$ of the intermediate gauge groups and the
SU(16) gauge group such that
                          \begin{eqnarray}
{\rm Tr}\, (T_iT_j)={1\over 2}\delta_{ij} \,, \quad
B_i={1\over f}\left(
{11\over3}N-{f\over3}(2n_g)-{1\over6}T(S_i)\right).\label{Bint}
                       \end{eqnarray}
Here, $f$ is the number of fundamentals of the subgroup per generation
in the 16 of Eq\@. (\ref{PsiL}). To be explicit, for the $SU(2)_{qL}$
group $f=3$,  for the SU(3)$_{qL}$ group $f=2$, and for all other
intermediate groups and the SU(16) group $f=1$.   The above equations
for $B_i$ hold for SU(N) gauge groups and with $N=0$ hold for U(1)
gauge groups. In the second terms, $n_g$, the number of fermion
generations, is multiplied with a factor of two to account for mirror
fermions.   This term does not affect the scales of symmetry
breaking.   In the third terms,  $T(S_i)$ is the quadradic Casimir
invariant for all Higgs submultiplets with masses less than the scale
of interest.   For a complex field,  the value of $T(S_i)$ should be
doubled.

With the above normalizations, the U(1) generators that
enter our symmetry breaking
pattern (see Fig\@. \ref{generalbreak}) are as follows:
                           \begin{eqnarray}
Q_{qB} &=& {1\over 2\sqrt{6}}\, \rm{diag}\,
           (0_{(4)},1_{(6)},-1_{(6)})  \,, \\
Q_{q\Lambda} &=& {1\over 2\sqrt{3}}\, \rm{diag}\,
          (0_{(4)},0_{(6)},1_{(3)},-1_{(3)}) \,, \\
Q_{qY} &=& {1\over 2\sqrt{33}}\, \rm{diag}\,
          (0_{(4)},1_{(6)},-4_{(3)},2_{(3)}) \,, \\
Q_{lX} &=& {1\over 2\sqrt{6}}\, \rm{diag}\,
          (1,1,1,-3,0_{(12)}) \,, \\
Q_{lY} &=& {1\over 2\sqrt{3}}\, \rm{diag}\,
          (0,-1,-1,2,0_{(12)}) \,, \\
Q_{Y} &=& {1\over 2\sqrt{15}}\, \rm{diag}\,
          (0,-3,-3,6,1_{(6)},-4_{(3)},2_{(3)}) \, .
                         \end{eqnarray}
In these equations, notation of the form $a_{(b)}$ stands for $b$
successive entries  of  $a$. For example, ``$1_{(6)}$'' stands
for ``1,1,1,1,1,1.'' The ordering is  the same as in Eq\@.
(\ref{PsiL}).

The breaking scheme shown in Fig\@. \ref{generalbreak}
has  the following boundary conditions at the
breaking scales $M_i$:
                        \begin{eqnarray}
\alpha^{-1}_{G}(M_{G}) = \alpha^{-1}_{12q}(M_{G}) &=&
     \alpha^{-1}_{4l}(M_{G})\,  ;\\
\alpha^{-1}_{12q}(M_{12}) = \alpha^{-1}_{6qL}(M_{12})
 &=& \alpha^{-1}_{6qR}(M_{12})=\alpha^{-1}_{1qB}(M_{12})\,  ;\\
 \alpha^{-1}_{6qL}(M_{6L}) =  \alpha^{-1}_{3qL}(M_{6L})
 &=& \alpha^{-1}_{2qL}(M_{6L})\,   ;\\
\alpha^{-1}_{6qR}(M_{6R}) = \alpha^{-1}_{3uR}(M_{6R})
 &=& \alpha^{-1}_{3dR}(M_{6R})=\alpha^{-1}_{1\Lambda q}(M_{6R})\,  ;\\
 {1\over2}\alpha^{-1}_{3uR}(M_{B})+{1\over2}\alpha^{-1}_{3dR}(M_{B})
&=& \alpha^{-1}_{3qR}(M_{B})\,  ,   \nonumber \\
{2\over11}\alpha^{-1}_{1qB}(M_{B})+{9\over11}\alpha^{-1}_{1\Lambda
q}(M_{B}) &=& \alpha^{-1}_{1qY}(M_{B})\,  ;\\
\alpha^{-1}_{4l}(M_{4l})
&=& \alpha^{-1}_{3l}(M_{4l})=\alpha^{-1}_{1lX}(M_{4l})\,  ;\\
 \alpha^{-1}_{3l}(M_{3l}) &=& \alpha^{-1}_{2lL}(M_{3l})\,   ,
  \nonumber \\
{1\over9}\alpha^{-1}_{3l}(M_{3l})+{8\over9}\alpha^{-1}_{1lX}(M_{3l})
&=& \alpha^{-1}_{1lY}(M_{3l})\,  ;\label{3l}\\
{2}\alpha^{-1}_{3Lq}(M_{Y})+{2}\alpha^{-1}_{3qR}(M_{Y})
&=& \alpha^{-1}_{3c}(M_{Y})\,   ,   \nonumber \\
{3}\alpha^{-1}_{2qL}(M_{Y})+\alpha^{-1}_{2lL}(M_{Y})
&=&\alpha^{-1}_{2L}(M_{Y})\,   ,   \nonumber \\
{11\over5}\alpha^{-1}_{1qY}(M_{Y})+{9\over5}\alpha^{-1}_{1lY}(M_{Y})
&=&\alpha^{-1}_{1Y}(M_{Y})\,   .\label{Y}
                      \end{eqnarray}
In writing Eq\@. (\ref{3l}), we have used the fact that  in the group
SU(3)$_l$ there is a generator
                        \begin{eqnarray}
\lambda_{3l}={1\over2\sqrt{3}}\, \rm{diag}\, (2,-1,-1,0,0_{(12)})
                      \end{eqnarray}
which is broken at the scale
$M_{3l}$. At  the same scale $U(1)_{lX}$ also breaks, leaving unbroken
the combination
                        \begin{eqnarray}
Q_{lY}= {1\over 3}\lambda_{3l} -{2\sqrt{2} \over 3}Q_{lX}\,.
                      \end{eqnarray}
  In deriving Eq\@. (\ref{Y}) for the breakings at the scale $M_Y$, we
have used similar considerations. In particular, SU(3)$_{qL}$, which
has the two diagonal generators
                          \begin{eqnarray}
\lambda_{3qL} &=& {1\over 2\sqrt{2}}\,
\rm{diag}\, (0_{(4)},1,-1,0,1,-1,0,0_{(6)}) \,, \\
\Lambda_{3qL} &= & {1\over
2\sqrt{6}}\, \rm{diag}\, (0_{(4)},-1,-1,2,-1,-1,2,0_{(6)}) \,,
                      \end{eqnarray}
 breaks, as does SU(3)$_{qR}$, which has the two diagonal generators
                        \begin{eqnarray}
\lambda_{3qR} &=& {1\over 2\sqrt{2}}\, \rm{diag}\,
(0_{(4)},0_{(6)},-1,1,0,-1,1,0)  \,, \\
\Lambda_{3qR} &=& {1\over
2\sqrt{6}}\, \rm{diag}\, (0_{(4)},0_{(6)},1,1,-2,1,1,-2) \,,
                      \end{eqnarray}
 leaving
unbroken SU(3)$_{c}$ which has the two diagonal generators
                        \begin{eqnarray}
\lambda_{3c} &=&\sqrt{2}\lambda_{3qL}+\sqrt{2}\lambda_{3qR} \,,\\
\Lambda_{3c} &=& \sqrt{2}\Lambda_{3qL}+\sqrt{2}\Lambda_{3qR} \,.
                      \end{eqnarray}
Similarly, $SU(2)_{qL}$ with the diagonal generator
                        \begin{eqnarray}
\lambda_{2qL}={1\over
2\sqrt{3}}\, \rm{diag}\, (0_{(4)},1_{(3)},-1_{(3)},0_{6}) ,
                      \end{eqnarray}
 and
$SU(2)_{lL}$ with the diagonal generator
                        \begin{eqnarray}
\lambda_{2lL}={1\over 2}\,
\rm{diag}\, (0,1,-1,0,0_{(12)})
                      \end{eqnarray}
 are broken, leaving $SU(2)_L$
unbroken with the diagonal generator
                        \begin{eqnarray}
\lambda_{2L}=\sqrt{3}\lambda_{2qL}+\lambda_{2lL}.
                      \end{eqnarray}

Now, using
                          \begin{eqnarray}
 n_i\equiv \log_{10} \left( {M_i\over {\rm 1 GeV}} \right) \, ,
                          \end{eqnarray}
the one-loop equations and boundary
conditions give us the following equations for the standard model couplings:
                         \begin{eqnarray}
\alpha^{-1}_{3c}(M_Z) = 4\alpha^{-1}_G &-&{\ln10\over2\pi}
      [(n_Y-n_Z)B_{3c} +(n_{3l}-n_Y)(2B_{3qL}+2B_{3qR})  \nonumber \\
&& +(n_{4l}-n_{3l})(2B_{3qL}+2B_{3qR})
+(n_B-n_{4l}) (2B_{3qL}+2B_{3qR})  \nonumber \\
&& +(n_{6R}-n_B)(2B_{3qL}+B_{3uR}+B_{3dR}) \nonumber \\
&& +(n_{6L}-n_{6R})(2B_{3qL}+2B_{6qR}) \nonumber \\
&&+(n_{12}-n_{6L})(2B_{6qL}+2B_{6qR}) \nonumber \\
&& +(n_G-n_{12})4B_{12q}]\,   ,   \\
\alpha^{-1}_{2L}(M_Z) = 4\alpha^{-1}_G &-&
 {\ln 10\over2\pi}[(n_Y-n_Z)B_{2L}
+(n_{3l}-n_Y)(3B_{2qL}+B_{2lL})  \nonumber \\
&&+(n_{4l}-n_{3l})(3B_{2qL}+B_{3l})
+(n_B-n_{4l})(3B_{2qL}+B_{4l})  \nonumber \\
&& +(n_{6R}-n_B)(3B_{2qL}+B_{4l})
+(n_{6L}-n_{6R})(3B_{2qL}+B_{4l})  \nonumber \\
&& +(n_{12}-n_{6L})(3B_{6qL}+B_{4l})  \nonumber \\
&& +(n_G-n_{12})(3B_{12qL}+B_{4l})]\,   ,
\\
\alpha^{-1}_{1Y}(M_Z)=4\alpha^{-1}_G &-& {\ln10\over2\pi}
  [(n_Y-n_Z)B_{1Y} +(n_{3l}-n_Y)({11\over5}B_{1qY}+{9\over5}B_{1iY})
\nonumber \\  &&+(n_{4l}-n_{3l})({11\over5}B_{1qY}+
          {1\over5}B_{3l}+{8\over5}B_{1lX})  \nonumber \\
&&+(n_B-n_{4l})({11\over5}B_{1qY}+ {9\over5}B_{4l})  \nonumber \\  &&
+(n_{6R}-n_B)({2\over5}B_{1qB}+{9\over5}B_{1\Lambda
q}+{9\over5}B_{4l})  \nonumber \\
&&+(n_{6L}-n_{6R})({2\over5}B_{1qB}+
  {9\over5}B_{6qR}+{9\over5}B_{4l})  \nonumber \\  &&
+(n_{12}-n_{6L})({2\over5}B_{1qB}+{9\over5}B_{6qR}+{9\over5}B_{4l})
\nonumber \\  && +(n_G-n_{12})({11\over5}B_{12q}+{9\over5}B_{4l})]\,
{}.
                                 \end{eqnarray}

For the couplings at $M_Z$ we use the experimental values
\cite{alphas}
                          \begin{eqnarray}
\alpha^{-1}_{3c}(M_Z) &=& 8.197\,   ,\nonumber  \\
\alpha^{-1}_{2L}(M_Z) &=& 30.102\,   ,\nonumber  \\
\alpha^{-1}_{1Y}(M_Z) &=& 59.217\,   ,\nonumber  \\
M_Z &=& 91.   176\,   {\rm GeV}\,   .
                        \end{eqnarray}
The $B_i$'s are determined by Eq\@. (\ref{Bsm}) and Eq\@.
(\ref{Bint})  with the $T(S_i)$'s being determined by the Higgs
structure given in Fig\@. \ref{generalbreak} along with the principal
of minimal fine-tuning. We note that to impliment the suggestion of
ref. \cite{unifiable} an additional rank-2 antisymmetric tensor is to
be included. However, this Higgs field has little effect on the RGE's
and is not included in the analysis of this section.

Now, the above equations for the standard model couplings can be solved
silmutaneously in terms of $n_G$ and $n_{12}$ to obtain
                      \begin{eqnarray}
n_G &=&
 - 3.28 - 0.09n_{Y} - 0.21n_{3l} + 0.90n_{4l}\nonumber \\  &&
 + 0.32n_B + 2.99n_{6R} - 2.60n_{6L}\,   ,  \\ n_{12} &=&
 - 2.08 - 0.02n_Y - 0.22n_{3l} + 0.73n_{4l}\nonumber \\  &&
 + 0.37n_B + 2.70n_{6R} - 2.34n_{6L}\,   .\label{generalsoln}
                       \end{eqnarray}
We note that Higgs fields make significant contributions to the above
equations. From the structure of the above equations, we see that
$n_{4l}$ and $n_{6R}$ being relatively high helps to meet the
requirement $n_G\geq   n_{12}$. In fact, no solution with low $n_{4l}$
is found to be acceptable.

We are further interested in seeing that it is possible to have new
physics, including breaking the Voloshin symmetry, at less than
1\,TeV.   Since our lowest intermediate stage, which contains the
Voloshin symmetry, is broken at $M_{3l}$,  we are interested in
$n_{3l}\alt 3$.   So as to investigate this possibility,  we make the
simplification
                         \begin{eqnarray}
M_{6L} & = & M_{12} = M_{G}\,   ,\nonumber  \\
M_{4l} & = & M_{B}= M_{6R}\, .\label{1}
                         \end{eqnarray}
This yields
                         \begin{eqnarray}
n_G &=& 9.35 + 0.66n_Y - 0.28n_{3l}\,   ,  \\
n_{6R} &=& 8.77 + 0.59n_Y - 0.19n_{3l}\,   .
                         \end{eqnarray}
In this example both $M_Y$ and $M_{3l}$ may be low. The particular
case of this  solution which has $M_Y=M_{3l}$ is graphed in Fig\@.
\ref{graphn6R} for the allowed region where $M_{G}\geq M_{12}\geq
M_{3l} \agt 250$ GeV. This gives the range of unification scale to be
$10^{10}\, {\rm GeV}\leq M_G\leq 10^{15}\, {\rm GeV}$.

We also investigate another possible solution where
                      \begin{eqnarray}
M_{4l} & = & M_{B}= M_{6R}= M_{6L}= M_{12}\,    ,\nonumber\\
M_{Y} & = &  10^{2} \, {\rm GeV} \,   .\label{2}
                      \end{eqnarray}
This yields
                      \begin{eqnarray}
n_G &=& 3.99+0.56n_{3l}\,   ,  \\
n_{12} &=& 4.62+0.48n_{3l}\,   .
                       \end{eqnarray}
We graph this case in Fig\@. \ref{graphn12}. Note that $M_G$ can be as
low as  $10^{8.5}$ GeV. In fact, from the constraint $M_{12} \geq
M_{3l}$, we note from  Fig\@. \ref{graphn12} that the unification
scale has to be smaller than $10^{8.9}$ GeV.
One characteristic of this
solution is that all intermediate scales other than $M_Y$ are larger
than $10^8$ GeV, so that the only observable new physics comes from the
scale $M_Y$.  However, we shall see in the next section that this
solution is inconsistent with the constraints arising from
nonobservation of proton decay, unless the discrete symmetry
$V^{\alpha} \rightarrow \, -V^{\alpha}$ is imposed on the Lagrangian.

Each of the solutions discussed above has the feature that one can

achieve grand unification at scales much lower than what is expected
in standard unification models based on gauge groups SU(5) or SO(10).
Such low
grandunification scale has many interesting features not present in
standard unification models. For example,
it has been shown \cite{monopole} that, unlike the SU(5) model,
one can circumvent the cosmological monopole problem easily. Also, the
intermediate scales are low, and some of them can be in reach of the
next generation of experiments. If new gauge structure exists at TeV
energies, as occurs for example in the second solution with low $M_Y$,
there will be new gauge bosons to mediate a lot of processes
\cite{Pal91}. Direct production of these gauge bosons at the SSC will
also provide exciting physics, some of which has been discussed in
the context of SU(15) models \cite{su15,rizzo}.

\section{Proton decay}\label{s:pdk}
            Since baryon and lepton numbers are part of the gauge
symmetry of the model, proton decay diagrams must involve baryon
number and lepton number violating
vacuum expectation values (VEV's) in the Higgs sector. The only
VEV in our symmetry breaking scheme which violates baryon number is
the VEV of the 560-dimensional representation, $B^{klm}$, having the
quantum numbers of $\left< \hat{u} \hat{d} \hat{d}
\right>$, which has $B=-1$. Lepton number violation arises from the
VEV of $B^{klm}$  having the quantum numbers of  $\left<
\hat{d}\left( ue-d\nu_e\right)\right>$, which has $L=1$, and the
VEV of 16-dimensional representation, $V^{\alpha}$, having the quantum
numbers of $\left< \hat{\nu}_e\right>$, which has $L=-1$. Note
that lowering indices changes the signs of the quantum numbers.

To examine proton decay, we use the method of effective operators. In
this method, one writes down all effective operators involving
fermions and scalars which are invariant under the full SU(16) group.
When the scalars develop VEV's, one obtains an effective operator
involving fermions only. These VEV's are responsible for baryon or
lepton violation.

The lowest order effective operators for proton decay involve four
fermionic fields \cite{protondecay}. From the discussion above, we
find two SU(16)-invariant effective operators which can induce proton
decay \cite{otherops}. These are
                       \begin{eqnarray}
{\cal O}_1 &=& \{ \Psi^k \Psi^l\} \{ \Psi^m \Psi^n\}
               B_{klr}B^{pqr}\Phi_{mp}\Phi_{nq}\, ,\\
{\cal O}_2 &=& \{ \Psi^k \Psi^l\} \{ \Psi^m \Psi^n\} B_{klm}V_{n}\, .
                       \end{eqnarray}
Here, $\{ \Psi^k \Psi^l\}\equiv \left( \Psi^k \right)^T C
\Psi_l$, where $C$ is the conjugation matrix for fermions.

Typical diagrams generating ${\cal O}_1$ and
${\cal O}_2$ are shown in Fig\@.
\ref{pdecayB}  and Fig\@. \ref{pdecayV}, respectively. In Fig\@.
\ref{pdecayB}, $\Phi_{u\hat{u}}$ gets a VEV.  In determining the
RGE's,  we only gave a VEV to the $\Phi^{e^-e^+}$ component of
{\bf 136}. However, in order to give masses to the quarks as well as
the charged leptons a VEV must be given to each of the
$\Phi^{u\hat{u}}$,  $\Phi^{d\hat{d}}$, and $\Phi^{e^-e^+}$
components of {\bf 136}, i.e. a linear combination of
$\Phi^{u\hat{u}}$,  $\Phi^{d\hat{d}}$, and $\Phi^{e^-e^+}$
represents the standard model Higgs. At higher scales  additional
components make contributions to the RGE's. These contributions are
small compared to other Higgs contributions. Therefore, we ignore
them.

{}From the figures we obtain an order-of-magnitude estimate of the
coeficients $\kappa_1$ and $\kappa_2$ of the 4-fermion operators ${\cal O}_1$
and ${\cal O}_2$  respectively:
                          \begin{eqnarray}
\kappa_1 &\sim & \left( {m_f\over
    M_W}\right)^2{\lambda_{B\Phi}\lambda_{BB}M_YM_BM_W^2\over M_G^6}\,
,\\ \kappa_2 &\sim & \left( {m_f\over
M_W}\right)^2 {\lambda M_BM_{3l}\over M_G^4}\, .
                         \end{eqnarray}
Here, the quantity $m_f$ is the mass of a typical fermion, and comes
from the Yukawa couplings. Antisymmetry of $B_{klr}$ and fermion
indices require use of second generation   fermions
\cite{protondecay}. Therefore, we use $m_f\simeq 100\, {\rm MeV}$.
Also, $\lambda_{B\Phi}$, $\lambda_{BB}$, and $\lambda$ denote the
scalar couplings. We have assumed that all virtual colored scalars have
masses of order $M_G$, the largest scale of this model. The mass
scales in the numerator are the scales of the VEV's. Making a rough
estimate, we have neglected factors of gauge coupling constants.

Known bounds on proton lifetime imply
                     \begin{eqnarray}
\kappa_1\, ,\, \kappa_2\alt 10^{-30} \, {\rm GeV}^{-2}\,.
\label{proton}
                      \end{eqnarray}
If we use the above constraint with $M_B\sim M_G$, which can be seen
to be true from Section \ref{s:rge}, and assume $\lambda_{B\Phi}$,
$\lambda_{BB}$, and $\lambda$ are $\sim 1$, then we find the
constraints
                     \begin{eqnarray}
{M_G^5 \over M_Y} &\agt & 10^{28}\, {\rm GeV}^4\, ,\label{O1}\\
{M_G^3 \over M_{3l}} &\agt & 10^{24}\, {\rm GeV}^2 \,. \label{O2}
                        \end{eqnarray}
from ${\cal O}_1$ and ${\cal O}_2$ respectively.  Eq\@. (\ref{O1}) puts no
restriction on the solutions to the RGE's, but Eq\@. (\ref{O2}),
resulting from ${\cal O}_2$, does. For example, Eq\@. (\ref{O2}) rules out
the entire special case defined by Eq\@. (\ref{2}) which gives $M_G$
as low as $\sim 10^{8.5}\, {\rm GeV}$, although it does not rule out
any region of the case defined by Eq\@. (\ref{1}) which allows for a
low energy $M_{3l}$. However, the effective operator ${\cal O}_2$, which
yields this constraint, is no longer allowed if we impose the discrete
symmetry $V^{\alpha}\rightarrow \, -V^{\alpha}$ on the Lagrangian.
Another feature that would exist if this discrete symmetry is imposed
is mentioned in the next section.

It is important to note that the decay modes of the proton
obtained from the operators ${\cal O}_1$ and ${\cal O}_2$ are different from
the
predictions of standard SU(5) or SO(10) grandunification models.
In ${\cal O}_1$, the indices $k,l$  are  antisymmetrized and so are
$m,n$. Thus, the quark level operator for proton decay
\cite{protondecay} arising from it is $\hat{u}\hat{s}\hat{u}\mu^+$.
This gives rise to the decay mode $p\rightarrow K^0 \mu^+$. On the
other hand, in ${\cal O}_2$, the quark level operator is
$\hat{u}\hat{s}\hat{d}\hat{\nu}$, where the neutrino can belong to any
generation since the indices $m,n$ are  not necessarily antisymmetric
for this operator. Thus, we expect a decay mode $p\rightarrow K^+
\hat\nu$. Note that both operators give rise to $B-L$ conserving
decays.

\section{Neutrino magnetic moment}\label{s:numu}
                           We now discuss generation of   a sizeable
magnetic moment for the neutrino.   The most general Yukawa couplings
of the model are given by
                         \begin{eqnarray}
-{\cal L}_Y  =  \sum_{a} h_a
\Psi_{aL}^\alpha \Psi_{aL}^\beta \Phi_{\alpha \beta}  +
  \sum_{a \not= b}
f_{ab}\Psi_{aL}^\alpha \Psi_{bL}^\beta
\varphi_{\alpha \beta} + \mbox{h.c.}
    \label{Yuk}
                         \end{eqnarray}
Here,  Latin indices refer to the generation.   $\Phi$ is the
symmetric 136-dimensional rank-2 SU(16) tensor representation,
whose couplings in the generation space are chosen diagonal without
loss of generality.   The additional multiplet $\varphi$ is a
120-dimensional antisymmetric rank-2 tensor,   so its coupling
$f_{ab}$ is antisymmetric in its generation indices. This field is put
into the model to generate a magnetic moment for the neutrino.  Since
quarks play no role in the magnetic moment of the neutrino,   we focus
our attention on the leptonic part of the interactions.

$\Phi$ is the Higgs field which breaks SU(3)$_c \times
{\rm SU(2)}_L \times {\rm U(1)}_Y$ down to
SU(3)$_c \times {\rm U(1)}_Q$.    In the leptonic sector we
assume only $\Phi_{34}$ gets a VEV
\cite{unifiable}.    This VEV gives masses to charged leptons,   but
not to neutrinos.    The multiplet $\varphi$,   on the other hand,
is assumed to have no VEV \cite{unifiable}.    The dominant
contributions to the mass of $\nu_e$ then come from the one-loop
graphs of Fig\@. \ref{diagram}.    The diagrams involve a Higgs
potential term of the form $\gamma \varphi^{\alpha \beta}
\varphi^{\gamma \delta} A_{\alpha \beta \gamma \delta}$,   where
$\gamma$ is a coupling with the dimension of mass and $A_{\alpha \beta
\gamma \delta}$ is the antisymmetric rank-4 SU(16) tensor whose VEV
$\left< A_{1234}\right>$ breaks SU(16) to $SU(12)_q \times SU(4)_l$
at $M_G$.   Because
 $A_{\alpha \beta \gamma \delta}$ is antisymmetric,  $\left<
A_{2314}\right>=-\left< A_{2413}\right>$, and therefore the  mass
contributions of the two diagrams of Fig\@. \ref{diagram} have
opposite sign.

{}From Fig\@. \ref{diagram},   we  estimate
                               \begin{eqnarray}
m_{\nu_e} \approx {f_{13}^2 \gamma \left< A_{2413}\right> m_\tau
\over 16 \pi^2 M^2_{24}} \ln\left( {M^2_{24} M^2_{13} \over
M^2_{14} M^2_{23}}\right)\,   ,
                                \end{eqnarray}
where $M_{\alpha \beta} \equiv (\rm{mass}\,  \rm{of}\,
\varphi_{\alpha \beta})$.    The important point to realize here is
that,   in the limit of unbroken SU(3)$_l$,   $M_{23}=M_{13}$ and
$M_{24}=M_{14}$,   and so the mass contributions from the two diagrams
cancel each other.

The diagrams of Fig\@. \ref{diagram} with one photon line attached
give the dominant diagrams for the magnetic moment of $\nu_e$.  These
diagrams add because of an extra negative sign between the two
diagrams arising from the photon vertex.   We  estimate
                        \begin{eqnarray}
\mu_{\nu_e} \approx {{e f^2_{13} \gamma \left< A_{2413}\right>
m_\tau}\over {16 \pi^2}}\,
 \left[ {1\over {M^2_{24} M^2_{13}}} + {1\over {M^2_{14} M^2_{23}}}
\right]\,   .
                        \end{eqnarray}

In the limit of unbroken SU(3)$_l$, we can write $M_{13}^2=M_{23}^2
\equiv M^2-{1\over 2}\Delta M_{4l}^2$ and $M_{14}^2=M_{24}^2 \equiv
M^2+{1\over 2}\Delta M_{4l}^2$, where $\Delta M_{4l}^2$  is the mass
difference due to SU(4)$_l$ breaking. The breaking of SU(3)$_l$ then
changes the masses according to
                        \begin{eqnarray}
M_{13}^2=M^2-{1\over 2}\Delta M_{4l}^2+\Delta M_{3l}^2 \, ,\,
M_{23}^2=M^2-{1\over 2}\Delta M_{4l}^2\, ,\label{3} \\
M_{14}^2=M^2+{1\over 2}\Delta M_{4l}^2+\Delta M_{3l}^2 \, ,\,
M_{24}^2=M^2+{1\over 2}\Delta M_{4l}^2\, , \label{4}
                        \end{eqnarray}
where $\Delta M_{3l}^2$ to lowest order is due to the term  $\lambda
\left< V_1\right> \left< V^1 \right> \varphi^{1\alpha}
\varphi_{1\alpha}$. Here, $V_\alpha$ is the vector representation
Higgs field and $\lambda$ is a dimensionless coupling. Putting  Eq\@.
(\ref{3}) and Eq\@. (\ref{4})  into the expressions for $m_{\nu_e}$
and  $\mu_{\nu_e}$ gives us
                        \begin{eqnarray}
m_{\nu_e} &\approx & {f_{13}^2 \gamma M_G m_\tau \over 16 \pi^2
M^6} \Delta M_{3l}^2 \Delta M_{4l}^2 \left[ 1-\left( {\Delta
M_{4l}^2 \over 2 M^2}\right)^2 \right]^{-1} \, ,\label{mass} \\
\mu_{\nu_e} &\approx & {e f_{13}^2 \gamma M_G m_\tau \over 8 \pi^2
M^4} \left[ 1-\left( {\Delta
M_{4l}^2 \over 2 M^2}\right)^2 \right]^{-1}\, ,\label{magmom}
                             \end{eqnarray}
where we have used $\left< A_{2413} \right> \sim M_G $ and
assumed    ${\Delta M_{3l}^2 \over M^2} \ll 1$. Now, requiring
$m_{\nu_e}\alt 10eV$ and $\mu_{\nu_e}\agt 10^{-11}\mu_B$, gives
from Eq\@. (\ref{mass}) and Eq\@. (\ref{magmom}) the constraints
                             \begin{eqnarray}
{f_{13}^2 \gamma M_G \over M^6}\Delta M_{3l}^2
\Delta M_{4l}^2 \left[ 1-\left(
{\Delta M_{4l}^2 \over 2 M^2}\right)^2 \right]^{-1}
&\alt & 10^{-6}\, ,\label{x} \\
{f_{13}^2 \gamma M_G \over M^4}
\left[ 1-\left( {\Delta M_{4l}^2 \over 2 M^2}\right)^2
\right]^{-1} &\agt & 10^{-6} \, {\rm GeV}^{-2}\, .\label{xx}
                             \end{eqnarray}
Also, demanding the mass-mixing matrix for $\varphi$ to have a
positive determinant gives the constraint
                               \begin{eqnarray}
\gamma < {M^2 \over M_G}\left[ 1-\left( {\Delta M_{4l}^2 \over
2 M^2 }\right)^2 \right]^{1\over 2}\, .\label{xxx}
                              \end{eqnarray}
We remark that if we impose the discrete symmetry $V_\alpha
\rightarrow -V_\alpha$ on the Lagrangian, then we should expect
$\gamma$ to be small because then $\gamma \varphi^{\alpha \beta}
\varphi^{\gamma \delta}A_{\alpha \beta \gamma \delta}$ is the only
term of the Lagrangian not invariant under  $A_{\alpha \beta \gamma
\delta}\rightarrow  -A_{\alpha \beta \gamma \delta}$. An example of
this is when in each of Eq\@. (\ref{x}), Eq\@. (\ref{xx}) and Eq\@.
(\ref{xxx}) the left and right sides are approximately equal. In this
case,  $\Delta M_{3l}^2 \Delta M_{4l}^2 \sim 10^6 f_{13}^2\, {\rm
GeV}^4$  and if also  $\left( {{\Delta M_{4l}^2} \over 2M^2} \right)^2
\ll 1$ then $M \sim 10^3 \, {\rm GeV}$. Typically, M is of the order
of 1\,TeV with SU(3)$_l$ and SU(4)$_l$ breaking inducing  small mass
changes in the $\varphi_{\alpha \beta}$'s with $\Delta M_{3l}^2 \sim
\Delta M_{4l}^2 \sim 10^3 \, {\rm GeV}^2$.

\section{Summary}
             We have shown in this paper  breaking chains for the
SU(16) grandunification group which lead to the standard model with
the quark and lepton sectors transforming separately at intermediate
energies. The grandunification scale for this model can be as low as
 $\sim 10^{10}\, {\rm GeV}$. We have shown that this does not produce
any conflict with know bounds of proton lifetime. In fact, if a
discrete  symmetry is imposed on the model, one can obtain chains with
the unification scale as low as  $\sim 10^{8.5} \, {\rm GeV}$. Also,
low intermediate breaking scales can
exist in the $\alt 1$\,TeV range. This has many observable
consequences, including guage bosons at the TeV range which can give
rise to a rich phenomenology.

Further, the model embeds the Voloshin symmetry $SU(2)_{\nu}$ into its
subgroup SU(4)$_l$. By using a rank-2 antisymmetric Higgs field, it is
possible to get a significant magnetic moment for a neutrino with
small mass. For this to work, this Higgs field typically should have
masses $\sim 1$\,TeV with relatively small mass differences induced by
SU(3)$_l$ and SU(4)$_l$ breaking.

This work has been supported by the Department of Energy grant
DE-FG06-85ER-40224.

\figure{{A possible chain of symmetry breaking. The numbers $n$ denote
a factor SU$(n)$ in the gauge group if $n>1$, and a U$(1)$ factor if
$n=1$. The superscripts $q$ or $l$ indicate whether only quarks (and
antiquarks) or only leptons (and antileptons) are non-singlets under
that part of the gauge group. If one considers the {\bf 255} as a
traceless matrix, its VEVs are diagonal and the notation {\bf
$1_{(6)}$}, e.g., stands for six consecutive entries of unity. In the
{\bf 560}, the symbol $\left< widehat{d} ue\right>$, e.g., stands for
the VEV of the color singlet combination of the components with one
index having the quantum numbers of $\hat{d}$, another of $u$ and
another of $e$. One can contemplate chains with fewer steps by
equating two or more energy scales. Note the following transformation
properties: $A^{[ijkl]}$~: {\bf 1820}, $T^k_l$~: {\bf 255},
$H^{[kl]}_{[pq]}$~: {\bf 14144}, $B^{[ijk]}$~: {\bf 560}, $V^i$~: {\bf
16}, $\Phi^{\{ ij\} }$~: {\bf  136}. }\label{generalbreak}}

\figure{{$n_G\equiv \log_{10}{M_G\over \, {\rm GeV}}$ and $n_{6R}
\equiv \log_{10}{M_{6R}\over \, {\rm GeV}}$ as a function of
$n_{3l}\equiv \log_{10}{M_{3l}\over \, {\rm GeV}}$ with $M_Y =
M_{3l}$, $M_{4l} = M_B = M_{6R}$ and $M_{6L}= M_{12}= M_G$ in  Fig.1.
The shaded area is not acceptable since by definition  $M_G\geq
M_{6R}\geq M_{3l}$.}\label{graphn6R}}.

\figure{{$n_G\equiv \log_{10}{M_G\over \, {\rm GeV}}$ and $n_{12}\equiv
\log_{10}{M_{12}\over \, {\rm GeV}}$ as a function of $n_{3l}\equiv
\log_{10}{M_{3l}\over \, {\rm GeV}}$ with $M_{4l} = M_B = M_{6R}=
M_{6L}= M_{12}$ and  $M_Y = 10^{2}\, {\rm GeV}$ in Fig. 1. The shaded
area is not  acceptable since by definition  $M_G\geq M_{12}\geq
M_{3l}$.}\label{graphn12}}.

\figure{{Tree diagram generating ${\cal O}_1$. The labels on the Higgs boson
lines represent, via Eq\@. (\ref{PsiL}), the transformation
properties under SU(16). The indices should all be considered as
upper indices.}\label{pdecayB}}

\figure{{Tree diagram generating ${\cal O}_2$.}\label{pdecayV}}

\figure{{The diagrams that contribute to the mass of the
neutrinos at the 1-loop level. Magnetic moment arises by
attaching a photon line to any internal line.} \label{diagram}}


\begin{thebibliography}{[000]}

\bibitem[*]{newadd} Present address: Center for
Particle Physics, Physics Department, University of Texas, Austin, TX
78712.

\bibitem{Adler} S.~L. Adler, Phys. Lett. {\bf 225B}, 143
(1989).

\bibitem{su15} P. H. Frampton and B-H. Lee,  Phys.
Rev. Lett. {\bf 64}, 619 (1990).

\bibitem{Pal91} P. B. Pal, Phys. Rev. {\bf D43}, 236 (1991).

\bibitem{su15higgs} B. Brahmachari, U. Sarkar, R.~B. Mann and T.~G.
Steele, Phys. Rev. {\bf D45}, 2467 (1992).

\bibitem{protondecay} P.~B. Pal, Phys. Rev. {\bf D45}, 2566 (1992).

\bibitem{ununified} S. Rajpoot, Mod. Phys. Lett. {\bf 1}, 645 (1986);
H. Georgi, E. Jenkins and E.~H. Simmons,
Phys. Rev. Lett. {\bf 62}, 2789 (1989); {\bf 63}, 1540 (E)
(1989); D. Choudhury, Mod. Phys. Lett. {\bf A6}, 1185 (1991).

\bibitem{su16} J.~C. Pati, A. Salam and J. Strathdee,
Nuovo Cim. {\bf 26A}, 77 (1975); Nucl. Phys. {\bf B185}, 445
(1981); R.~N. Mohapatra and M. Popovi\'c, Phys. Rev. {\bf
D25}, 3012 (1982); A. Raychaudhuri and U. Sarkar, Phys. Rev.
{\bf D26}, 3212 (1982).

\bibitem{Volo} M.~B. Voloshin, Sov. J. Nucl. Phys. {\bf 48},
 512, (1988).

\bibitem{unifiable} N.~G. Deshpande and P.~B. Pal,  Phys. Rev. {\bf
D45}, 3183 (1992).

\bibitem{Davis} R. Davis et. al., Phys. Rev. Lett. {\bf 20}, 1205
(1968); J.~K. Rowley, B.~T. Cleveland and R. Davis: {\em Solar
Neutrinos and Neutrino Astrophysics}, AIP Conference Proceedings No.
126 (American Institute of Physics), p. 1.

\bibitem{anticorr} G.~A. Bazilevskaya, Yu.~I. Stozhkov and
T.~N. Charakhch'yan, JETP Letters 35, 341 (1982). For detailed
references and discussion, see, e. g. , J.~N. Bahcall and W.~H.
Press, Astrophys. J. {\bf 370}, 730 (1991).

\bibitem{monopole} P.~B. Pal, Phys. Rev. {\bf D44}, R1366 (1991).

\bibitem{otherpd} P.~H. Frampton and T.~W. Kephart, Phys. Rev. {\bf
D42}, 3892 (1990).

\bibitem{BarbMoha} R. Barbieri and R.~N. Mohapatra, Phys. Lett.
{\bf 218B}, 225 (1989).

\bibitem{others} K.~S. Babu and R.~N. Mohapatra, Phys.
Rev. Lett. {\bf 63}, 228 (1989); K.~S. Babu and R.~N. Mohapatra,
Phys. Rev. Lett. {\bf 64}, 1705 (1990); G. Ecker, W. Grimus
and H. Neufeld, Phys. Lett. {\bf B232}, 217 (1990); D. Chang,
W.~Y. Keung and G. Senjanovi\'c, Phys. Rev. {\bf D42}, 1599
(1990); N. Marcus and M. Leurer, Phys. Lett. {\bf B237},
 81 (1990); H. Georgi and L. Randall, Phys. Lett. {\bf 244B},
 196 (1990).

\bibitem{VVO} A. Cisneros, Astrophys. Space Sc. {\bf 10}, 87,
(1971); M. Voloshin, M. Vysotskii and L.~B. Okun, JETP
{\bf 64},  446, (1986); Sov. J. Nucl. Phys. {\bf 44}, 440,
(1986).

\bibitem{finetune} F. del Aguila and L. Iba\~{n}ez, Nucl. Phys. {\bf
B177}, 60 (1981); R.~N. Mohapatra and G. Senjanovi\'{c}, Phys. Rev.
{\bf D27}, 1601 (1983).

\bibitem{alphas} The central values of $\alpha_1$ and $\alpha_2$ are
taken from U. Amaldi, W. de Boer and H. F\"urstenau, Phys. Lett. {\bf
B260}, 447 (1991). The central value of $\alpha_3$ is taken from T.
Hebbecker: plenary talk at the Lepton-Photon symposium, Geneva, July
1991.

\bibitem{rizzo}
T.~G. Rizzo, Phys. Rev. {\bf D46}, 910 (1992).

\bibitem{otherops} One can easily construct other operators which
violate baryon and lepton numbers, but they do not contribute to
proton decay. For example, consider
$\{ \Psi^k \Psi^l\} \{ \Psi^m \Psi^n\}  B_{klr}B^{pqr} A_{mnpq}$. If
only  VEVs are those shown in Fig. \ref{generalbreak},
it is easy to see that no four-fermion operator results.

\end{thebibliography}
\end {document}